\documentstyle[12pt]{article}
\advance\topmargin by -3\baselineskip
\advance\textheight by 4\baselineskip
\begin{document}

\newcommand{\be}{\begin{equation}}
\newcommand{\ee}{\end{equation}}
\newcommand{\bea}{\begin{eqnarray}}
\newcommand{\eea}{\end{eqnarray}}
\newcommand{\btab}{\begin{tabular}}
\newcommand{\etab}{\end{tabular}}
\newcommand{\bef}{\begin{figure}}
\newcommand{\eef}{\end{figure}}
\newcommand{\bt}{\begin{table}}
\newcommand{\et}{\end{table}}
\newcommand{\ben}{\begin{enumerate}}
\newcommand{\een}{\end{enumerate}}
\newcommand{\ba}{\begin{array}}
\newcommand{\ea}{\end{array}}
\newcommand{\twowaypartial}{i \!\!\stackrel{\leftrightarrow}{\partial}}
\newcommand{\half}{\frac{1}{2}}
\newcommand{\simleq}{\stackrel{<}{\sim}}
\newcommand{\simgeq}{\stackrel{>}{\sim}}
\newcommand{\slashp}{\not \!p}
\newcommand{\slashq}{\not \!q}
\newcommand{\slashk}{\not \!k}
\newcommand{\slashepsilon}{\not \!\epsilon}
\newcommand{\slashP}{\not \!\!\!\:P}
\newcommand{\slashQ}{\not \!\!\!\:Q}
\newcommand{\slashK}{\not \!\!\!\:K}

\thispagestyle{empty}
\centerline{\begin{tabular}{c}
{\Huge  Beyond The Colour-Singlet Model For }\\
  {\Huge Inelastic $J/\psi$ Photoproduction}\\
\\
\\
\\
\\
{\bf Hafsa Khan}\\
Department of Physics    \\
Quaid-e-Azam University  \\
Islamabad, Pakistan. \\
and    \\
{\bf Pervez Hoodbhoy}   \\
Center for Theoretical Physics  \\
Massachussets Institute of Technology   \\
Cambridge, MA 02139, USA   \\
and  \\
Department of Physics    \\
Quaid-e-Azam University  \\
Islamabad, Pakistan.
\end{tabular}
\pagebreak                                             }

\begin{abstract}
Bound-state corrections to $J/\psi$ production from almost real photons are
calculated in the colour-singlet model. A systematic, gauge-invariant, theory
 of hard quarkonium processes is used upto $O(v^2)$, where $v$ is the relative
velocity of the quarks. The internal structure
of the meson is characterised by two parameters, $\epsilon_B/M$ and
$\nabla^2 \phi(0)/M^2 \phi(0)$, in addition to the usual wavefunction at the
origin $\phi(0)$. These parameters are constrained to be consistent with
measured
leptonic decay of the $J/\psi$ and hadronic and radiative decays of $\eta_c$.
The calculated corrections to the colour-singlet model, which include
radiative effects, improve agreement with the experimental data.
\end{abstract}

\newpage

The main production mechanism for $J/\psi$ particles in the inelastic process
$\gamma+g \longrightarrow J/\psi+X$ is believed to be the fusion,
$\gamma+g \longrightarrow J/\psi+g$. Because of this, $J/\psi$ production is an
important tool for exploring the gluonic distribution inside nucleons. The
first calculation of this in the so-called ``colour-singlet" model was
performed by Berger and Jones\cite{Berger} over 15 years ago. Subsequently,
several
authors applied the model to data as they became available. Recently radiative
corrections to the basic model were calculated by Kr\"{a}mer et
al.\cite{Zerwas},
who
found these to be large at moderate photon energies $E_\gamma \approx 100$GeV.
Earlier, relativistic corrections had been estimated by Jung et al.\cite{Jung}
using a
model proposed by Keung and Muzinich\cite{Keung}. These authors found the
corrections to be fairly substantial, especially in the high-z $(z \ge 0.8)$
region where the validity of the colour-singlet model is suspect. Inelastic
photoproduction of the $J/\psi$ has been reviewed by Ali\cite{Ali}.

The purpose of this letter is to explore the effect of the binding of
the quarks upon $J/\psi$ photoproduction in the colour-singlet model. In
the original calculations\cite{Berger} this was totally
neglected and the $c,\bar{c}$ were put on their mass shells. This is a
sensible starting point because the binding energy $\epsilon_B=2 m\!-\!M$, and
the quark relative velocity $v$, are small parameters:
$\epsilon_B/M \ll 1$ and $v^2/c^2 \ll 1$. Since the $c$ quark mass is only
$\sim 1.5$ GeV, substantial corrections could exist. This is indeed suggested
by the calculations of refs.\cite{Jung,Keung}. However, we do not find the
calculational method convincing for two important reasons. First, the model
of ref\cite{Keung} does not treat
gauge-invariance satisfactorily. Second, it incorporates binding energy
corrections but not wavefunction corrections. Additionally, systematic
improvement of the model seems difficult. Therefore a re-examination of
bound-state corrections is important.

We have recently developed a systematically improvable gauge-invariant
formalism for the one and two photon (gluon) decays of heavy quarkonia
\cite{Hafsa}. We
extend and apply this here to $\gamma+g \longrightarrow J/\psi+g$. This
involves three bosons and is therefore considerably more complicated in
computational terms. As noted in \cite{Hafsa}, the
$J/\psi$ internal structure is described by more parameters than simply $\phi
(0)$, the
quark wavefunction at zero separation. These are
$\epsilon_B/M$ and $\nabla^2 \phi(0)/M^2 \phi(0)$. Our starting point is
that the photoproduction amplitude
$\gamma+p \longrightarrow J/\psi+X$ is given by the sum of all distinct Feynman
diagrams leading from the initial to the final state (Fig.1a). Each diagram
can be written as an integral over the loop momenta which, for the lowest order
diagram illustrated in Fig.1b, is
\be
T^{\mu_1 \mu_2 \mu_3}_{o(1b)} = \int \frac{d^4k}{(2 \pi)^4}
Tr \left[ M(k) H^{\mu_1 \mu_2 \mu_3} (k) \right].
\ee
The tensor $H^{\mu_1 \mu_2 \mu_3} (k)$ is the amplitude to produce a
free gluon and two quarks, not necessarily on their mass-shells, from
a photon and gluon. We call this
the ``hard" or perturbative part, and its expression can be read off from
Fig.1b and permutations. The ``soft" part is the zero-gluon, non
gauge-invariant,
Bethe-Salpeter amplitude,
\be
M(k) = \int d^4\! x \,e^{\iota k\cdot x} \langle 0 |T [\bar{\psi} (-x/2)
\psi(x/2) ] | P,\epsilon \rangle.
\ee
In equations 1-2, $x^{\mu}$ is the relative distance between quarks and
$k^{\mu}$ is the relative momentum.

The next category of diagrams contain a single gluon exchange between the
blob and one of the two hard propagators (Fig.2). These all have the general
form,
\be
T_1^{\mu_1 \mu_2 \mu_3} = \int \frac{d^4 k}{(2 \pi)^4}
\frac{d^4 k'}{(2 \pi)^4} Tr \,\, [M^{\rho}(k,k') H_{\rho}^{\mu_1 \mu_2 \mu_3}
(k,k')].
\ee
Again $H^{\mu_1 \mu_2 \mu_3}_\rho$ may be directly read off from the diagrams,
and the soft part is,
\be
M^{\rho}(k) = \int d^4 \!x \, d^4 \! z \,e^{\iota k\cdot x} e^{\iota k'\cdot z}
\langle 0 |T [\bar{\psi} (-x/2) A^{\rho}(z) \psi(x/2) ] | P,\epsilon \rangle.
\ee
This is a matrix in colour space since,
$A^{\rho} \!\! \equiv \!\! \frac{1}{2} \lambda^a A^{a \rho}$.
The soft gluon which originates
from the blob has its momentum $k'$ bounded by $R^{-1}\simleq k' \ll M$, where
$R$
is the meson's spatial size. The two gluon diagram can be included in the same
way,
\begin{equation}
T_{2 }^{\mu_1 \mu_2 \mu_3} = \int \frac{d^4k}{(2 \pi)^4} \frac{d^4k'}{(2
\pi)^4}
\frac{d^4k''}{(2 \pi)^4}  Tr  M^{\rho' \rho''}(k,k',k'') H^{\mu_1 \mu_2
\mu_3}_{\rho'
\rho''} (k,k',k''),
\end{equation}
with,
\be
M_{\rho' \rho''\!}(k,\!k'\!,\!k''\!)\!=\!\!\int \!\!d^4x d^4x'\!
d^4x''\!e^{i(k.x+k'\!.x'+k''\!.x'')}
         \! \langle 0 | T[\bar{\psi}(-x/2) A_{\rho'}\!(x'\!)
A_{\rho''}\!(x''\!) \psi (x/2)] | P,\epsilon \rangle.
\ee
The gluon self-interaction diagram in Fig.3b is similarly included.

As the next step, the hard parts are expanded in the quark and gluon relative
momenta. The various amplitudes are combined and the Ward identity
$\partial^{\alpha} S_F = - S_F \gamma^{\alpha} S_F$ is freely used. The upshot
of the calculations is that the ordinary derivatives combine with gauge
fields to yield covariant derivatives, i.e., a gauge-invariant result for the
amplitude for $\gamma+g \longrightarrow J/\psi+X$ is obtained,
\bea
\label{7}
(T_0+T_1+T_2)^{\mu_1 \mu_2 \mu_3} &=& Tr [ \langle 0 |\bar{\psi}\psi |
P,\epsilon
\rangle h^{\mu_1 \mu_2 \mu_3} \nonumber \\
&+& \langle 0 |\bar{\psi} i\!\! \stackrel{\leftrightarrow}{D}_{\alpha}\psi
|P,\epsilon
\rangle \partial^{\alpha} h^{\mu_1 \mu_2 \mu_3}  \nonumber \\
&+& \langle 0 |\bar{\psi} i\!\! \stackrel{\leftrightarrow}{D}_{\alpha}
i\!\! \stackrel{\leftrightarrow}{D}_{\beta}
\psi | P,\epsilon \rangle \frac{1}{2} \partial^{\alpha} \partial^{\beta}
h^{\mu_1 \mu_2 \mu_3} \nonumber \\
&+&\langle 0 |\bar{\psi} F^{\alpha \beta}\psi | P,\epsilon \rangle
\frac{i}{2} \partial'_{\alpha}H_{\beta}^{\mu_1 \mu_2 \mu_3}+ \dots ].
\eea
In the above, $h^{\mu_1 \mu_2 \mu_3}=H^{\mu_1 \mu_2 \mu_3}(k=0)$. The last
term shall not concern us here since it is of higher order than $v^2$.

To proceed, one can perform a Lorentz and CPT invariant decomposition of each
of the hadronic matrix elements in Eq.\ref{7}. This is somewhat complicated
\cite{MAli}
and involves a large number of constants which characterize the hadron.
Considerable simplification results from choosing the Coulomb gauge, together
with the counting rules of Lepage et al.\cite{Lepage}. The result of using this
analysis
is that, in this particular gauge, the gluons contribute
at $O(v^3)$ to the reaction $^3 S_1 \rightarrow \gamma + X$  and hence can
be ignored. Even this leaves us with too many parameters, and forces us to
search for a dynamical theory describing the essential dynamics of a $Q
\bar{Q}$
system going beyond the usual non-relativistic potential models. A possible,
but by no means unique, description is provided
by the Bethe-Salpeter equation with an instantaneous kernel. This has been
conveniently reviewed by Keung and Muzinich\cite{Keung} and we shall use
their expression for the B-S amplitude in terms of the non-relativistic
wavefunction
$\phi(p)$. By projecting appropriately from their wavefunction  it is readily
established
that for $1^{--}$ states,
\bea
\label{me}
\langle 0 | \bar{\psi} \psi |P,\epsilon\rangle & = & \frac{1}{2} M^{1/2} \!
\left( 1\!+\!\frac{\nabla^2}{M^2} \right) \phi \left( 1\!+\!\frac{\slashP}{M}
\right)
\slashepsilon-\frac{1}{2} M^{1/2} \! \frac{\nabla^2 \phi}{3 M^2} \left(
1\!-\!\frac{\slashP}{M} \right)  \slashepsilon, \nonumber \\
\langle 0 |\bar{\psi}  \twowaypartial_{\alpha}\! \psi | P,\epsilon \rangle  &=&
-\frac{1}{3} M^{3/2}\frac{\nabla^2\phi}{M^2}\epsilon^\beta
\left( -g_{\alpha \beta}+i \epsilon_{\mu \nu \alpha \beta} \frac{P^\nu}{M}
\gamma^\mu \gamma_5 \right) , \nonumber \\
\langle 0 |\bar{\psi}  \twowaypartial_{\alpha}  \twowaypartial_{\beta}\!\psi |
P,\epsilon
\rangle &=& \frac{1}{6}M^{5/2} \frac{\nabla^2 \phi}{M^2} \left( g_{\alpha
\beta}
-\frac{P_{\alpha}P_{\beta}}{M^2} \right)  \left( 1 \!+\! \frac{\slashP}{M}
\right) \slashepsilon.
\eea

Using Eqs.\ref{7} and \ref{me}, the differential crossection for the subprocess
$\gamma+g \longrightarrow J/\psi+g$ comes out to be \footnote{We used
Mathematica \cite{Math}, supplemented by the
HIP package \cite{HIP}, for computation of traces and simplification of
algebra.},
\be
\label{dsig}
\frac{d\sigma}{dt} = \frac{256}{3 s^2}\alpha_e \alpha_S^2 \pi^2 M e_q^2
                     |\phi(0)|^2 [\eta_0 f_0(s,t,u)+\eta_B f_B(s,t,u)+
                                  \eta_Wf_W(s,t,u)].
\ee
In the above $s$, $t$, and $u$ are the partonic level Mandelstam variables (we
omit
the usual carets) which obey the
relation $s+t+u=M^2$. The appropriate average over initial gluon colours
and sum over final gluon colours has been made. If radiative corrections are
ignored,
\be
\eta_0=1, \,\, \eta_B=\frac{\epsilon_B}{M}, \,\,
\eta_W=\frac{\nabla^2\phi}{M^2\phi}.
\ee
The function $f_0$ is the standard, leading order, result:
\be
f_0(s,t,u) = \frac{(s^2 t^2+t^2 u^2+u^2 s^2 +M^2 s\,t\,u)}{(s-M^2)^2 (t-M^2)^2
(u-M^2)^2}.
\ee
The binding energy and wavefunction corrections, $f_B$ and $f_W$ respectively,
are slightly more complicated:
\bea
f_B(s,t,u) &=& \frac{1}{4 D} \left[-7s\,t\,u(s^4+t^4+u^4)+7M^2(s^3 t^3+t^3 u^3
+u^3 s^3)\right.
\nonumber \\ &+&(s^2 t^2+ t^2 u^2 + u^2 s^2)(s^3+t^3+u^3+15s\,t\,u)
\nonumber \\ &+&\left. M^2s\,t\,u(s^3+t^3+u^3)+29 M^2 s^2 t^2 u^2 \right],
\eea
and,
\bea
f_W(s,t,u) &=& \frac{1}{6 D}\left[ 141s\,t\,u(s^4+t^4+u^4)-85 M^2(s^3 t^3+t^3
u^3 +u^3 s^3) \right.
\nonumber \\&-&27(s^2 t^2+ t^2 u^2 + u^2
s^2)(s^3+t^3+u^3+\frac{205}{27}s\,t\,u)
\nonumber \\&-&\left. 139 M^2 s\,t\,u(s^3+t^3+u^3) -463 M^2 s^2 t^2 u^2
\right].
\eea
The denominator $D$ is\footnote{We note that Eq.\ref{dsig}
reduces to Eq.23 of Jung et al.\cite{Jung} if the condition
$\eta_W=\frac{1}{2} \eta_B$ is imposed. This latter condition is equivalent
to $\frac{1}{M} \nabla^2 \phi(0)=\frac{1}{2} \epsilon_B \phi(0)$, which is the
Schr\"{o}dinger equation for quark relative motion in a potential which
vanishes at zero separation. It is also worthy of note that the same
condition emerges as a renormalization condition in the treatment of
positronium by Labelle et al\cite{Labelle} (see their Eqs.11 and 12). However,
in our treatment there is no principle which apriori constrains $\eta_B$ to
bear a fixed relation to $\eta_W$ and therefore both will be considered
adjustable parameters.},
\be
D = (s-M^2)^3 (t-M^2)^3 (u-M^2)^3.
\ee
The total $\gamma p$ crossection is obtained by convoluting $d\sigma/dt$
with the gluon distribution $G(x)$ in the proton.
\be
\label{conv}
\frac{d^2 \sigma}{dx dt} = G(x) \frac{d\sigma}{dt}.
\ee

Integration of Eq.\ref{dsig} over $t$ in the interval $M^2-s$ to $0$ yields,
\be
\label{sigma}
\sigma(s)=\frac{256}{3} \pi^2 \alpha_e \alpha_s^2 e_q^2
\frac{|\phi(0)|^2}{M^5} [\eta_0 F_0+\eta_B F_B+\eta_W F_W+rad. cor.],
\ee
where,
\bea
F_0 &=& \left[ -1-4\xi+2 \xi^3+\xi^4+2\xi^5-2\xi(1+2\xi+5\xi^2)\log \xi \right]
/
(1-\xi)^2 \xi^2
(1+\xi)^3,\nonumber \\
F_B &=&\frac{1}{2}\left[ -2+16\xi-10\xi^2+48\xi^3+10\xi^4-64\xi^5+2\xi^6
\right. \nonumber \\
&-& \left. \left( 1-3\xi+14\xi^2-106\xi^3+17\xi^4 -51\xi^5 \right) \log \xi
\right]/
(1-\xi)^3 \xi^2 (1+\xi)^4, \nonumber \\
{\rm and} \nonumber \\
F_W &=& \frac{1}{3} \left[ 26-14\xi+210\xi^2-134\xi^3-274\xi^4+150\xi^5+38\xi^6
-2 \xi^7 \right.\nonumber \\
&+&\!\!\!\!\! \left.
(27\!+\!50\xi\!+\!257\xi^2\!\!-\!292\xi^3\!+\!205\xi^4\!-\!78\xi^5\!-\!41\xi^6)\log\! \xi \right]\!/
(1\!-\!\xi)^3 \xi^2(1\!+\!\xi)^5.
\eea
In the above, $\xi=s/M^2$. The curves for $F_0$, $F_B$ and $F_W$ as a function
of
$\xi$ are shown in Fig[4]. Note that  $F_W$ is a large negative number for very
small values of $s$ which rises and becomes positive for $\sqrt{s}>4.75 \:GeV$
while
 $F_B$ is a negative quantity for all values of the incoming photon energies.
The radiative correction to $\sigma(s)$ has already been calculated by
Kr\"{a}mer et al.\cite{Zerwas}. They have
taken into account the modification of the initial gluon densities as well as
the box diagrams and the splitting of the final gluon into gluon and light
quark-antiquark pairs. Like the behaviour of the scaling functions
plotted in \cite{Zerwas}, the corrections presented in Fig.[4] are also large
at
moderate photon energies but decrease with incoming energies.

The differential crossection $d\sigma/dz$ calculated for $\sqrt{s}=14.7 GeV$
is shown in Fig.[5] where $z=E_{J/\psi}/E_\gamma$. We use a simple gluon
distribution
function $x G(x)=3 (1-x)^5$.
For the numerical evaluations we take $\alpha_s=0.19$ and $m=1.43 GeV$.
Additionally, we take $\eta_W=-.073$ and $|R_{J/\psi}|^2 =0.978 GeV^3$. This
set of parameters is the same as given in Ref.\cite{Hafsa}. These values,
when inputted into the theoretical formulae, with radiative corrections
evaluated at $\mu=m$ yield the following values for the decay
widths\cite{Hafsa},
\begin{eqnarray}
\Gamma (J/\psi \longrightarrow e^+ e^-) & = & 5.61 \;\;KeV \nonumber \\
\Gamma (\eta_c \longrightarrow hadrons) & = & 9.99 \;\;MeV  \nonumber \\
\Gamma (\eta_c \longrightarrow 2 \gamma) & = & 6.48 \;\;KeV
\end{eqnarray}
which are quite close to the values of the experimentally measured decay
widths\cite{Field} $5.36 \pm .28 \:KeV,\: 10.3 \pm 3.6 \:KeV$ and
$8.1 \pm 2 \: KeV$ respectively. As remarked in Ref.\cite{Hafsa}, the value of
$\alpha_s$ chosen for the numerical calculations differs from its value
deduced from deep inelastic scattering, $\alpha_s(m_c) \approx 0.3$.
Larger values inserted into the expressions for the decay rates result
in large differences between the wavefunctions at the origin of
$J/\psi$ and $\eta_c$, violating the assumption that these
are of $O(v^2)$.

In Fig.[5], we show various theoretical results for $d\sigma/dz$, together
with the experimental data taken from the EMC and NMC\cite{EMC1,EMC2,NMC}.
The dash-dot line is the zeroth order term calculated by
setting $\eta_B=\eta_W=0$ in Eq.[\ref{dsig}]. $d\sigma/dz$ calculated from
Eq.[\ref{dsig}] with the identification $\eta_W=\frac{1}{2} \eta_B$, but
without radiative corrections, is shown with a dotted line (see footnote 2).
The dashed line
is the differential crossection for $\eta_W=-.073$ without radiative
corrections.
Finally the solid line is the full result (including radiative
corrections) calculated upto $O(v^2)$. A $K$ factor of $3.5$ has been used,
obtained by fitting the crossection including all the corrections upto $O(v^2)$
to the experimental data\footnote{The value of the $K$ factor depends
upon the value of $\alpha_s$. For $\alpha_s=0.3$, it is lowered to 1.25. }.
The overall effect of including these terms is that
 they describe the experimental data quite well in the region $0.5<z<0.9$.

In Fig.[6a]-[6f], the double differential crossection $d\sigma/{dzdp_t^2}$
is shown for different $z$-bins and a comparison is made with the
experimental data \cite{EMC1,EMC2}. We see that the relativistic corrections
move
the theoretical predictions in the right direction. However
the theoretical curve does not describe the data for the highest $z$ region
$(0.95<z<1)$ where the elastic $J/\psi$ production is expected to be
important. Our formalism works best for the moderate $z$ region i.e,
$0.7 \leq z \leq 0.95$. Note that the validity of the colour singlet model
hinges essentially upon all propagators being hard, a condition which only
holds
away from the end points. Gluonic vacuum fluctuation, a non-perturbative
effect, may otherwise be important\cite{Vol}.

In conclusion, we have estimated the $O(v^2)$ corrections to the inelastic
$J/\psi$
production from $\gamma p$ collisions and shown that the agreement of theory
with
experiment is improved if these corrections are included.

\vspace{4.5cm}
\centerline{\bf Acknowledgements}
 We thank D.Wyler for identifying an important error in the preliminary
version of this paper, and for numerous comments. This work was supported in
part by funds provided by NSF Grant No. INT-9122027. H.K gratefully
acknowledges
financial support from the Akhter Ali Memorial Fund for her doctoral
research.

\newpage
\begin{center}
\section*{\bf Figure Captions}
\end{center}
Figure:1 a) The amplitude $\gamma+p \longrightarrow J/\psi+X$,
         b)  One of the 6 leading order diagrams.\\
Figure:2 a)   One gluon exchange diagram.\\
Figure:3 a)   Two gluon exchange diagram.
         b) Three gluon vertex diagram.\\
Figure:4   The functions $F_0$, $F_B$ and $F_W$ as a function of $\xi=s/M^2$.
Note
that their contribution is large at moderate incoming photon energies and
decreases as $s$ increases.\\
Figure:5   The differential crossection $d\sigma/dz$ plotted as a function of
z at $\sqrt{s}=14.7 GeV$.
The solid line is the full crossection
(including the radiative correction calculated by Kr\"{a}mer et
al.\cite{Zerwas}).
 $d\sigma/dz$ calculated from Eq.[9] with $\eta_B=-0.076$ and $\eta_W=-.073$ is
shown
by the dashed line while the dotted line corresponds to the curve with the
choice $\eta_W=\frac{1}{2}\eta_B$ (see footnote 2).
The curve represented by dash-dot line is
the crossection with $\eta_B=0$ and $\eta_W=0$. A $K$ factor of $3.5$ has
been used to account for the overall normalization. The data points are taken
from EMC and NMC\cite{EMC1,EMC2,NMC}. \\
Figure:6   The double differential crossection $d\sigma/dz/dp_t^2$ for
different z-bins.
 The solid curve is predicted from the model including $\eta_B$ and $\eta_W$
corrections ($K=3.5$). The dotted curve is for the choice
$\eta_W=\frac{1}{2}\eta_B$.
 The dashed curve is for $\eta_B=0$ and $\eta_W=0$.

\end{document}